\documentclass[twocolumn, longauthor]{aastex631}
\usepackage{academicons}
\usepackage{orcidlink}
\usepackage{array}
\definecolor{orcidlogocol}{HTML}{A6CE39}
\newcolumntype{P}[1]{>{\centering\arraybackslash}p{#1}}

\begin{document}

\title{Astrometric Binary Classification Via Artificial Neural Networks}

\author[0009-0008-3226-6205]{Joe Smith}
\affiliation{Boyce Research Initiatives and Education Foundation, San Diego, CA 92106, USA; \href{jsmithsd.ohs@gmail.com}{jsmithsd.ohs@gmail.com}}

\begin{abstract}

With nearly two billion stars observed and their corresponding astrometric parameters evaluated in the recent \textit{Gaia} mission, the number of astrometric binary candidates have risen significantly. Due to the surplus of astrometric data, the current computational methods employed to inspect these astrometric binary candidates are both computationally expensive and cannot be executed in a reasonable time frame. In light of this, a machine learning (ML) technique to automatically classify whether a set of stars belong to an astrometric binary pair via an artificial neural network (ANN) is proposed. Using data from \textit{Gaia} DR3, the ANN was trained and tested on 1.5 million highly probable true and visual binaries, considering the proper motions, parallaxes, and angular and physical separations as features. The ANN achieves high classification scores, with an accuracy of 99.3\%, a precision rate of 0.988, a recall rate of 0.991, and an AUC of 0.999, indicating that the utilized ML technique is a highly effective method for classifying astrometric binaries. Thus, the proposed ANN is a promising alternative to the existing methods for the classification of astrometric binaries.

\end{abstract}

\keywords
{\href{http://astrothesaurus.org/uat/293}{Computational astronomy (293)} --- \href{http://astrothesaurus.org/uat/79}{Astrometric binary stars (79)} --- \href{http://astrothesaurus.org/uat/1933}{Neural networks (1933)} ---\href{http://astrothesaurus.org/uat/1907}{Classification (1907)}}

\section{Introduction} \label{sec:intro}

\indent 
A binary star system is where two or more stars are gravitationally bound, orbiting about a common center of mass. Astrometric binaries are binary star systems inferred by considering the astrometric parameters (positions, proper motions, parallaxes, and radial velocities) of each component star. Astrometric binaries provide significant value, as other than being used to map the cosmos at large, their astrometric parameters can be used to determine the masses of the component stars (\citet{2020AJ....159...33Z}), examine open cluster and star field kinematics (\citet{2023A&A...675A.180G}), identify limits on the dark matter density present in massive compact halo objects (\citet{2004ApJ...601..311Y}), and can be utilized in general gravitational probes (\citet{2024MNRAS.528.4720H}). Hence, it is very important that binary star catalogs are constructed in order to be applied to the research of a variety of other astrophysical phenomena.

ML techniques, and deep neural networks (DNNs) in particular, have been applied to astronomy at large in recent years, prompting many new discoveries and questions (e.g., \citet{Lin_2020}; \citet{Li_2020}; \citet{szk22}). One especially popular method of DNNs, ANNs, have been commonly applied to problems involving signal analysis and random function approximations. ANNs have been extensively applied to classification problems pertaining to astronomy, such as the morphological classification of galaxies (e.g., \citet{1996MNRAS.283..207L}; \citet{2010MNRAS.406..342B}), the classification of galaxy spectra (\citet{1992MNRAS.259P...8S}; \citet{10.1093/mnras/283.2.651}), the classification of stellar spectra (\citet{1994MNRAS.269...97V}; \citet{10.1046/j.1365-8711.1998.01596.x}), and the selection of pulsar candidates from radio surveys (\citet{10.1111/j.1365-2966.2010.17082.x}). However, while astronomers continue to automate the classifications and predictions of other astrophysical phenomena, they have largely overlooked applying such ML techniques to astrometry. 

In recent times, sky surveys and space missions have obtained vast amounts of data on individual stars and their relative positions and motions. Namely, the \textit{Gaia} mission (\citet{2016A&A...595A...1G}) launched by the European Space Agency (ESA) has observed over 1.5 billion stars and determined their respective astrometric parameters with unprecedented precision. Due to the incredible scale of this mission, a surplus of new binary star candidates have been introduced for classification. In light of this, astronomers have employed computational techniques to compute binary catalogs by considering the astrometric parameters of potential binary star pairs (e.g., \citet{10.1093/mnras/stab323}; \citet{2023AJ....166..218M}). However, not only are these methods restricted to only limited subsets of \textit{Gaia} data, they are also inefficient at classifying a group of stars as binary, often requiring many intermediate steps before successfully classifying an input set of stars. Consequently, it has proven to be very difficult to analyze the entire \textit{Gaia} database to construct an appropriate binary catalog. Therefore, an automated technique for the classification of astrometric binaries is required.

In this paper, an ANN model is proposed for the automated classification of astrometric binaries. The proposed ANN aims to solve the problem posed by the present methods of the classification of astrometric binaries in order to maximize the potential of \textit{Gaia} data releases and future surveys. This work also offers an opportunity to compare human classifications to those from automated ML algorithms on a large scale. If proven to be as effective as human classifications of astrometric binaries, the proposed ANN could save significant computing time and costs for future surveys and studies.

The rest of this paper is organized as follows. In Section \ref{sec:data}, the data and the methods for data preprocessing are presented, while Section \ref{sec:ANN} describes the proposed ANN. The results and associated discussions are given in Section \ref{sec:results}. The paper closes with Section \ref{sec:conclusions}, in which the conclusions about the efficacy of the proposed ANN are given.

\section{Data and Methods} \label{sec:data}
Given the scale and precision of the \textit{Gaia} mission, this work will utilize data from \textit{Gaia} Data Release 3 (DR3; \citet{2023A&A...674A...1G}) to train and test the proposed ANN.

\subsection{Feature Selection} \label{subsec:feat}

In the construction of the data set, the radial velocities (RVs) of each binary are neglected. This is because of the more than 1.5 billion stars observed by the \textit{Gaia} mission, only 33 million have their RVs measured. As a result, the proposed ANN will only be trained on a subset of the overall astrometric parameters of each binary: the positions, proper motions ($\mu_1$, $\mu_2$ mas yr$^{-1}$), and parallaxes ($\varpi_1$, $\varpi_2$ mas). Here, $\mu_1$ and $\mu_2$, and $\varpi_1$, $\varpi_2$, indicate the proper motion and parallax values of the primary and secondary component stars, respectively. The declination ($\delta_1$, $\delta_2$) and the right ascension ($\alpha_1$, $\alpha_2$) of the component stars of each binary cannot be meaningful alone, but they can be used to compute their associated angular ($\theta$\textdegree) and physical (\textit{s} au) separations. {The angular separation between any two component stars is defined as} 

\begin{equation} \label{eq:asep}
    \frac{\Theta}{\arcsec} = 2\sin^{-1}(\sqrt{\sin^2\frac{\Delta\delta}{2} + \cos\delta_1\cos\delta_2\sin^2\frac{\Delta\alpha}{2}}),
\end{equation}
{where $\Delta\delta = \delta_1 - \delta_2$ and $\Delta\alpha = \alpha_1 - \alpha_2$. The angular separation is converted into degrees by $\theta = \frac{\Theta}{3600}$. The physical separation is computed as}
\begin{equation} \label{eq:psep}
    \frac{s}{\text{au}} = 1000\frac{\Theta}{\varpi},
\end{equation}
{serving as an accurate approximation despite not being equivalent to the complete 3D separation (\citet{2018ElBadry}).}

These six features (inputs) will be considered to train and test the ANN. The aim is to effectively train the ANN to map the features of binary star candidates to their correct class: true binaries (positives) or visual binaries (negatives).

\subsection{Numerical Data} \label{subsec:data}

Many binary star catalogs have been computed utilizing the data available from \textit{Gaia} data releases. One of the most extensive is provided in \citet{10.1093/mnras/stab323}, where 1.1 million highly probable binary star candidates are calculated within ~1 kpc of the Sun from \textit{Gaia} DR3. In deriving the catalog, conditions are imposed on certain astrometric parameters of each candidate component star: the parallax, proper motion, and physical separation values. Most notably, the parallax and proper motion values of potential binary star candidates are calculated such that they are within an appropriate relative range in comparison to their potential companion star. In other words, the catalog is derived on the basis of similar proper motion and parallax values. In regard to physical separations, the component stars must have a physical separation of less than 1 pc, thereby accounting for a range of binary star systems, such as wide binaries. This threshold is chosen since a very limited binary population is expected to exist at physical separations larger than this.

The catalog derived by El-Badry et al. is sifted for both highly probable true binaries and visual binaries, which are commonly referred to as “chance-alignments.” This is done by sifting the computed chance-alignment rates,\begin{em}R\end{em}, between 0.9 $<$\begin{em}R\end{em}$<$ 1.1, where\begin{em}R\end{em}\url{~} 1 is a highly probable visual binary, and\begin{em}R\end{em}$<$ 0.1, where\begin{em}R\end{em}\url{~} 0 is a highly probable true binary. This yielded 1.5 million highly probable true and visual binaries, where true binaries were then assigned a “binary class” value of “0” and visual binaries a binary class value of “1”. The parallaxes, proper motions, and physical and angular separations of each true and visual binary are then extracted into a separate data set. Since the selected features do not include the relative distances of each binary candidate from the Earth or the Sun, the selection criteria initially imposed will not restrict the applicability of the trained ANN to arbitrary relative distances.

\begin{figure}[t]
    \centering
    \includegraphics[width=1.09\linewidth]{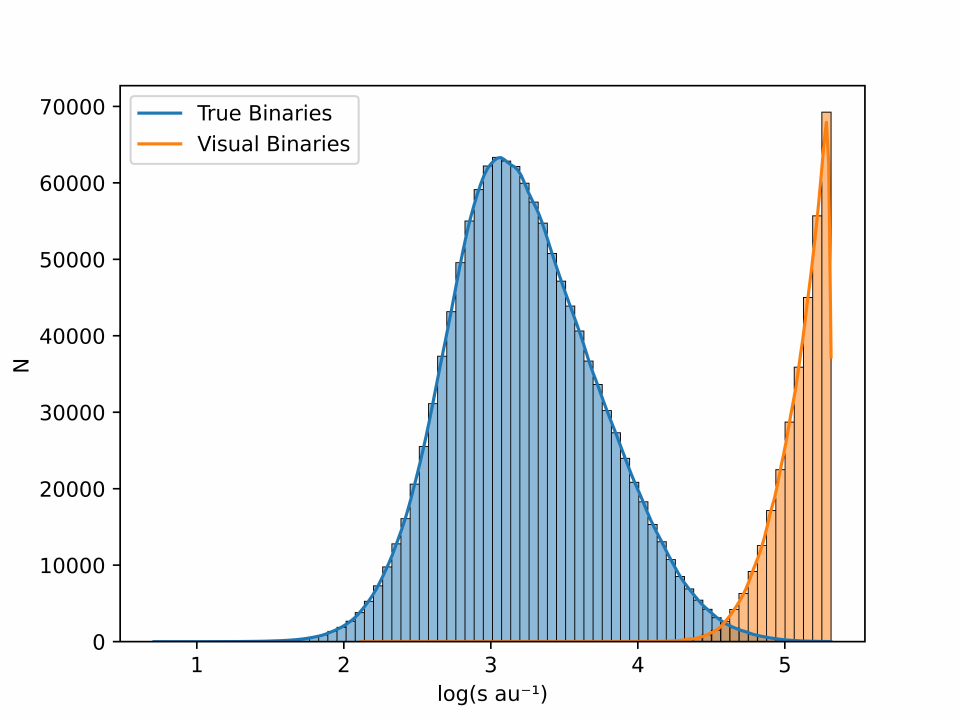}
    \caption{Histogram of true and visual binaries in the sorted catalogue as a function of their physical separation. The range of values fall between 0.70 $<$ log(\textit{s} au$^{-1}$) $<$ 5.31, obeying the initial condition of \textit{s} $<$ 206265 au.}
    \label{fig:physepnetdata}
\end{figure}

\begin{figure}[t]
    \centering
    \includegraphics[width=1.0\linewidth]{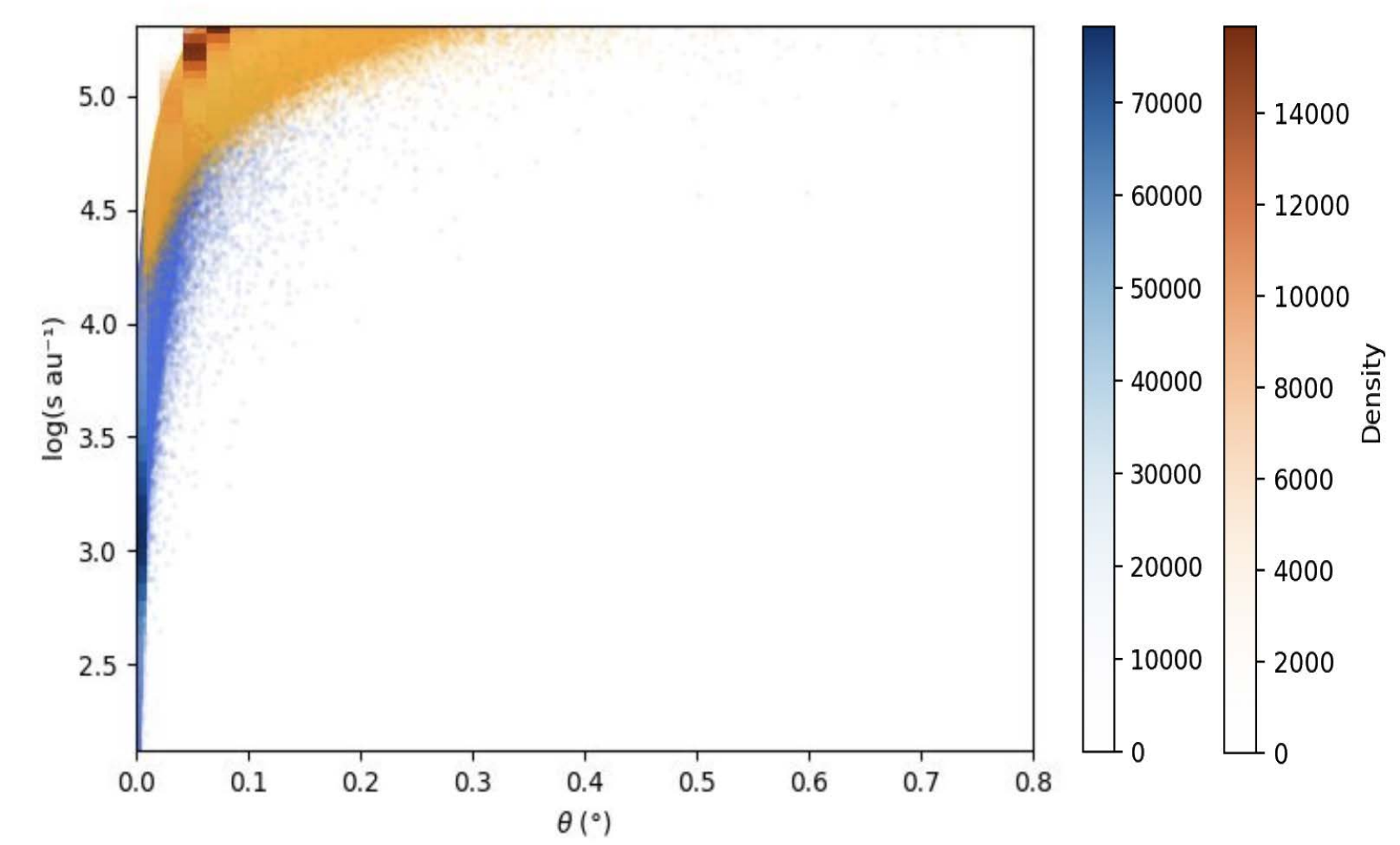}
    \caption{{Physical separations plotted against their corresponding angular separations. Overplotted for true (blue) and visual (orange) binaries are 2D density histograms.}}
    \label{fig:thetavs}
\end{figure}

\subsection{Class Imbalance} \label{subsec:imbalance}

Considering the range of potential values for each astrometric parameter associated with either a true or visual binary, it is expected that there will be some degree of class imbalance present in the data. Classification algorithms often suffer from imbalanced data due to the skew that is present towards one class. As a result, classification algorithms are easily affected by an imbalanced data set, which commonly leads to a poor classifier despite an increase in algorithm performance (\citet{2013arXiv1305.1707L}; \citet{article}). The level of class imbalance is measured as the ratio between true binaries to visual binaries, which is called the imbalance ratio (IR). {A popular standard defines a high class imbalance when the IR ranges from 100:1 to 10,000:1 (\citet{5128907}). There is an IR of 4:1 in the data set used in this work, indicating that the data set utilized will not significantly inhibit or inflate model classification scores considering the shape of the true and visual binary classes.}

{Figure \ref{fig:physepnetdata} shows a histogram of each true and visual binary as a function of their physical separation. As can be seen in this figure, the physical separation distributions of each class indicate that the data set is strongly defined by the physical separation feature. Comparing the physical separation feature against the angular separation feature in Figure \ref{fig:thetavs} yields a similar result, where there is similar clustering found in the equivalent physical separation regions. However, the binary classes are not as well defined by their angular separations as by their physical separations, where the true and visual binary classes have mean angular separations of 0.0023° ± 0.0060° and 0.065° ± 0.039°, respectively. The implications for ML model performance as a result of the strong class definition given by the physical separation feature will be explored in greater depth in Section \ref{sec:modelcomps}.}

\subsection{Data Preprocessing} \label{sec:preproc}
The data set (1.5 million binaries) is split into three parts: training set (60\%), validation set (20\%), and test set (20\%). The training set is used to fit the proposed ANN, while the validation set provides the information necessary to tune the selected hyperparameters of the ANN. Since there is limited class imbalance present in the data set, a data augmentation approach to minimize the effects of class imbalance during the training, validation, and testing phases is unnecessary. 

\section{Artificial Neural Network} \label{sec:ANN}
There are many ML techniques available to apply to the problem of classifying binary star systems. For this work, an ANN is chosen due to the very large numerical data set that is used, which is ideal for training DNNs compared to, for example, support vector machines (SVMs; \citet{708428}). Compared to other DNNs, an ANN is also likely to be preferred for the classification of astrometric binaries. {In particular, classification of via a convolutional neural network (CNN; \citet{lecun2015deep}) may not be feasible due to the limited information provided by images. However, additional work is necessary to confirm that CNNs cannot reliably classify astrometric binaries.}

\subsection{Architecture of the Artificial Neural Network} \label{sec:archit}
\begin{figure}
        \centering
        \includegraphics[width=1\linewidth]{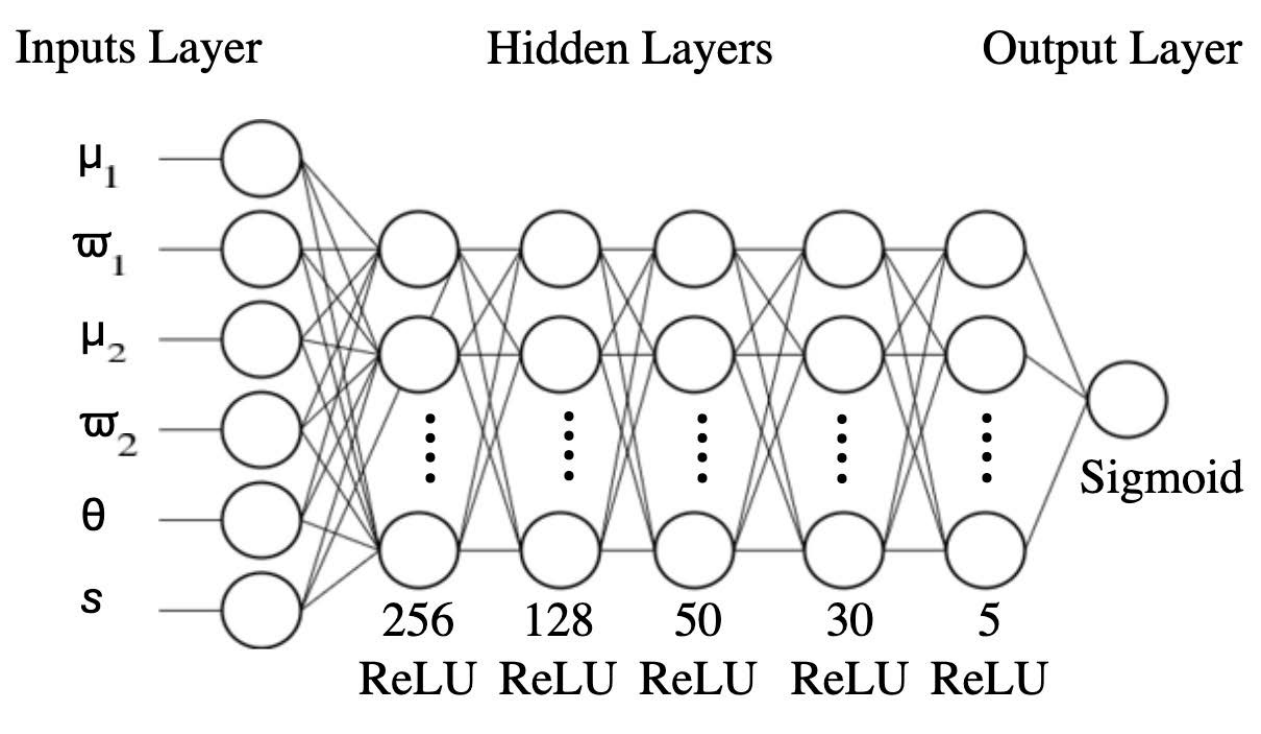}
         \caption{The architecture of the proposed ANN, where the inputs are the six features associated with the input binary candidate (Section \ref{subsec:feat}). See Table \ref{tab2} for the model architecture of the ANN in detail.}
    \label{fig:architpic}
\end{figure}

The proposed ANN was developed using the Keras API built within TenserFlow (\citet{tensorflow2015-whitepaper}), an open source library for ML and deep learning. The architecture of this ANN can be seen in Figure \ref{fig:architpic}. 

The astrometric parameters in the sorted catalogue are essentially floating point numbers, which can be used as input parameters in the ANN without requiring any manipulation. Therefore, a simple dense layer with a sigmoid activation was applied after the hidden (dense) layers resolved a numerical input via rectified linear unit activation functions (ReLU; \citet{10.5555/3104322.3104425}). This outputs a probability that determines whether or not the input binary star system is a visual or true binary under an operating threshold. The selected threshold for this ANN was the standard 0.5. If the output provided by the sigmoid activation function is greater than 0.5, then the input binary star system is classified as a visual binary, whereas if the output is less than 0.5, then the input binary star system is classified as a true binary. The classifications were compared with the ground truth of labeled classes using the binary cross entropy loss function (\citet{doi:10.1126/science.1127647}) and batch stochastic gradient descent (SGD) with the Adam optimizer (\citet{2014arXiv1412.6980K}).

{Cross-validation (CV) methods are typically used to minimize the classification error of the DNN by providing information on the required tuning of the model hyperparameters (\citet{hastie2009}; \citet{10.1214/09-SS054}). Of the many CV methods available, this work employs a method known as hold-out CV due to the nature of the objectives of this work. The potential for overfitting due to hold-out CV is minimized by optimizing the hyperparameters of the learning algorithm (\citet{pub.1035636638}).}

\subsection{Model Optimization} \label{sec:optim}

Hyperparameters of mini-batch size, number of epochs, and learning rate were optimized for more efficient model training. The model performed best when a relatively small mini-batch size of 512 and a Keras learning rate of 0.00025 were used. To prevent overfitting, a regularization step of early stopping was applied (\citet{NIPS2000_059fdcd9}). Via a callback of validation loss, the early stopping function monitors the change of the validation loss value before stopping the training process and saving the best model weights. The number of epochs without improvement after which training will be stopped by the early stopping function is referred to as "patience." In this work, a patience of 20 and a minimum required change of 10$^{-4}$ are utilized. Table \ref{tab2} presents the tested and optimal hyperparameters of the ANN. 

\begin{table}[t]
\setlength{\tabcolsep}{1.6pt}
\centering
\caption{Hyperparameters of the ANN model}
\renewcommand{\arraystretch}{1.3}
\scriptsize
\begin{tabular}{l*{5}cc}
\hline
\hline
$$Parameter$$ & $$Tested Values$$ & & $$Selected Value$$ \\
\hline
\colhead{\textit{Architecture}} \\
Number of dense layers & [1 -- 10] & & 5 \\
& $$ReLU$$ & & \\
Dense activation function & $$Sigmoid$$ & & $$ReLU$$ \\
& $$Tanh$$ & &\\
Number of neurons per layer & [1 -- 1000] & & $$Fig. $$ \ref{fig:architpic} \\
\hline
\colhead{\textit{Optimization}} \\
Mini-batch size & [16 -- 2048] & & 512 \\
Learning rate & [10$^{-2}$ -- 10$^{-5}$] & & \(2.5 \times 10^{-4}\) \\
Optimizer & & $$Adam$$ & \\
\hline
\end{tabular}
\label{tab2}
\end{table}

\begin{figure*}[hbt]
\centering
 \includegraphics[width=.49\linewidth]{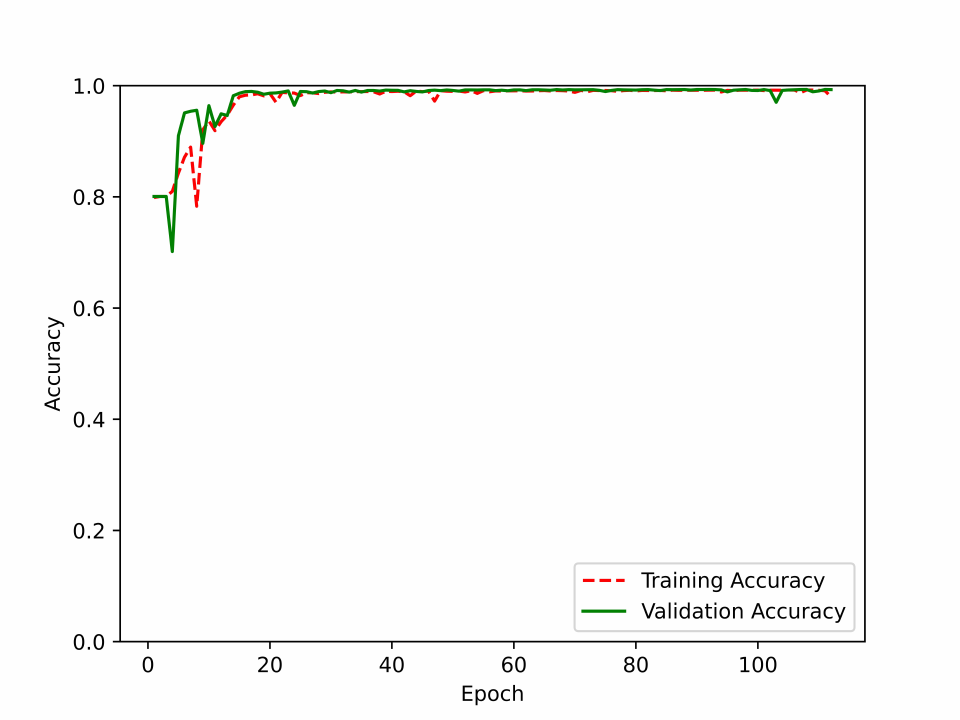}
 \includegraphics[width=.49\linewidth]{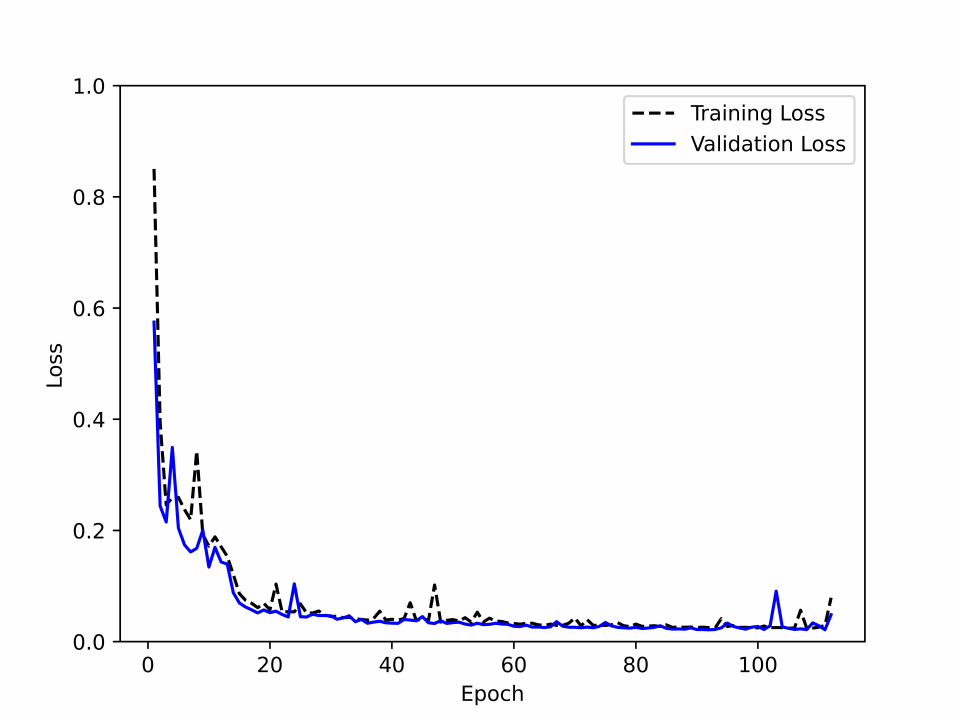}

 \includegraphics[width=.49\linewidth]{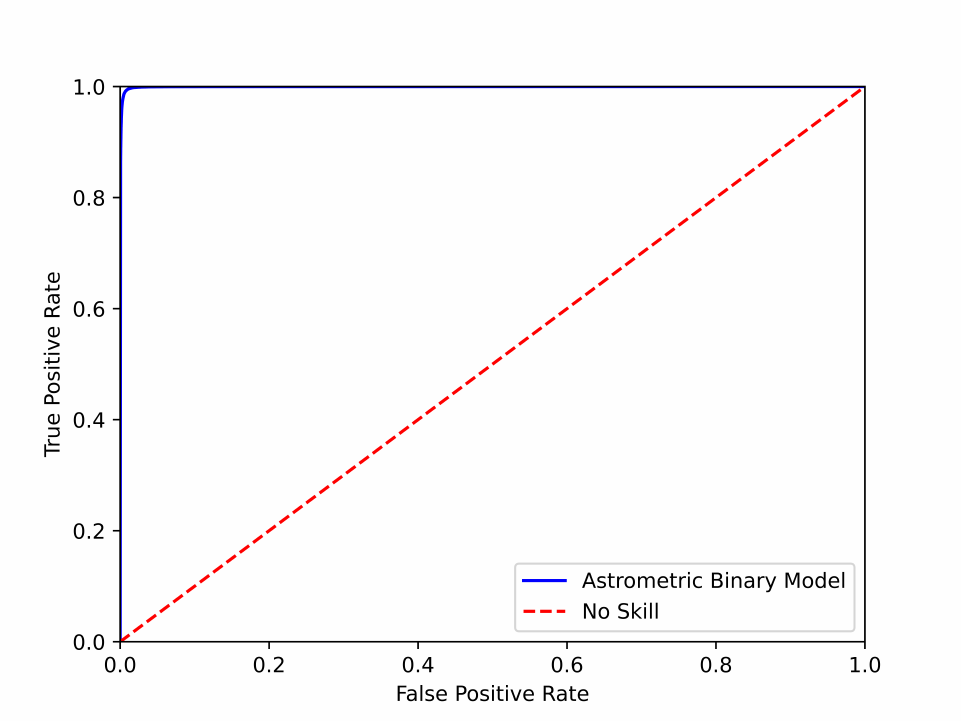}
 \includegraphics[width=.49\linewidth]{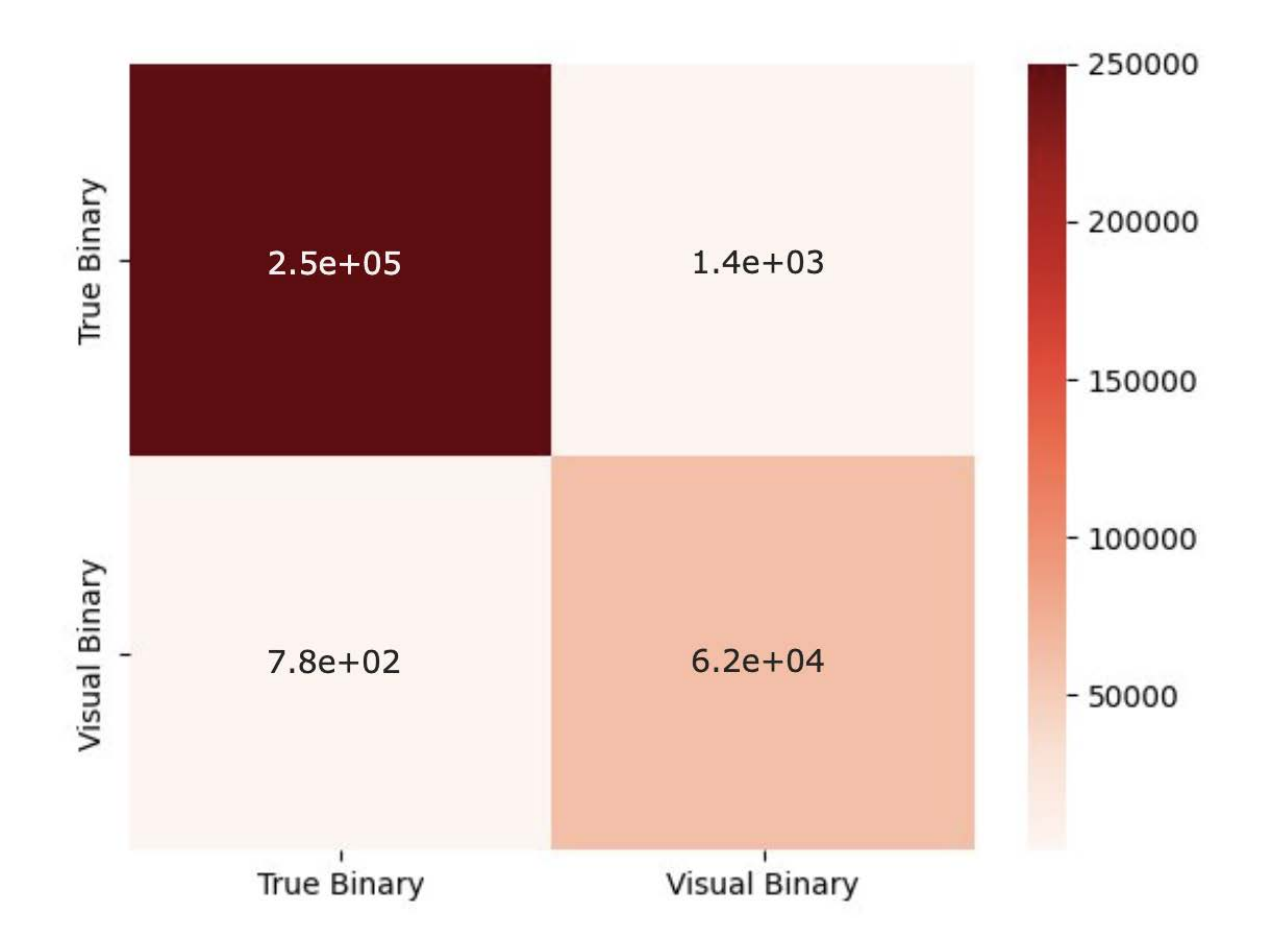}
\caption{The first row presents the performance of the ANN considering the accuracy (left) and loss (right) evolutions during the training and validation phases. The second row presents the computed ROC curve (left) and confusion matrix (right) during the test phase. \\}
\label{fig:perffig}
\end{figure*}

\subsection{Performance Measure} \label{sec:perfmeas}

In order to evaluate the performance of the DNN, several classification scores should be considered, such as the accuracy, the precision, the recall, and F1 score (\citet{10.1093/mnras/staa218}). However, the conventional practice of evaluating classification models via a singular evaluation metric, most often the accuracy score, does not often provide adequate information (\citet{5128907}). Consequently, the accuracy, recall, precision, and area under the curve (AUC) scores are each considered to evaluate the performance of the proposed ANN. 

With respect to the application of pattern recognition tasks to this work, accuracy measures how often the proposed ANN correctly classified the input binary star as either a visual binary or a true binary. Similarly, recall measures how many binaries would be correctly classified as true binaries from all the true binaries, and precision measures how many true binaries would be correctly classified from all true binary candidates. {Meanwhile, the AUC score indicates how well a model can discriminate, or rank, between random examples from the data set via their computed class probability across varying thresholds. To compute an AUC score, a receiver operating characteristic (ROC) curve is necessary. The ROC curve is created by plotting the true positive rate (TPR) as a function of the false positive rate (FPR). The FPR is defined as the probability that a true binary label is given to an input visual binary while the TPR is defined as the probability that a true binary label is given to an input true binary. When a learning algorithm has an AUC of 0.5, it is considered that the model has a random classification scheme (“no skill").} Each of these metrics have a score range of 0 to 1, with 1 representing perfect model classification.

\section{Results and Discussion} \label{sec:results}

In the following section, the training, CV, and testing performances are each evaluated via the scikit-learn open source ML library (\citet{scikit-learn}). {The misclassified binaries are also investigated and solutions are proposed for improving model performance. Furthermore, the influence on the data set given by the physical separation feature is investigated by constructing ML comparison models trained on a subset of the features of the proposed ANN. With current knowledge, this work is the first to develop an automatic classification tool for astrometric binary star systems.}

\subsection{Performance and Analysis} \label{sec:perfanaly}

The evolution processes of the accuracy and loss scores as a function of epoch are shown in Figure \ref{fig:perffig}. The ANN converges very quickly, reaching high accuracy and low loss scores within 30 epochs. This can be attributed to the optimization of the selection of hyperparameters and the large data set that was used to train and validate the ANN. Besides occasional deviations, the accuracy function continually increased and the loss function continually decreased before the early stopping regularization step halted the training process within 112 epochs. To evaluate model performance, a ROC curve (\citet{roccurve}) and a confusion matrix (\citet{Visaref}) are computed and shown in Figure \ref{fig:perffig}. 

\begin{figure}[t]
    \centering
    \includegraphics[width=1.05\linewidth]{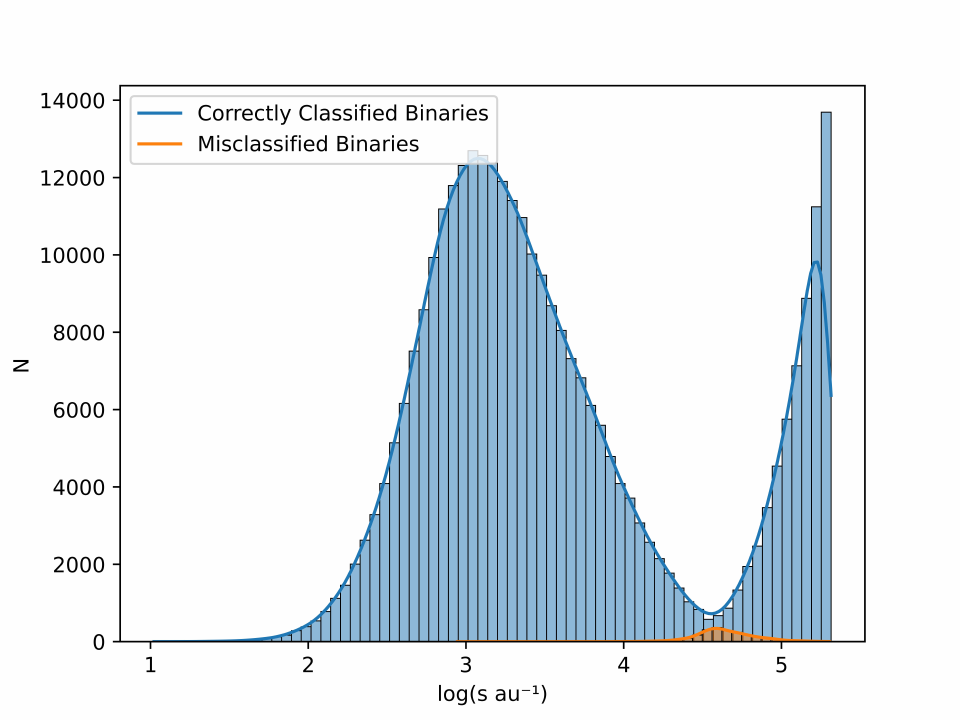}
    \caption{Histogram of misclassified and correctly classified binaries from the test set as a function of their physical separation. The range of physical separations for misclassified binaries fall between 2.95 $<$ log(\textit{s} au$^{-1}$) $<$ 5.30, with values falling between 4.10 $<$ log(\textit{s} au$^{-1}$) $<$ 5.18 within ±$3\sigma$ of the mean.}
    \label{fig:misclasscomp}
\end{figure}
\begin{figure}[t]
    \centering
    \includegraphics[width=1.05\linewidth]{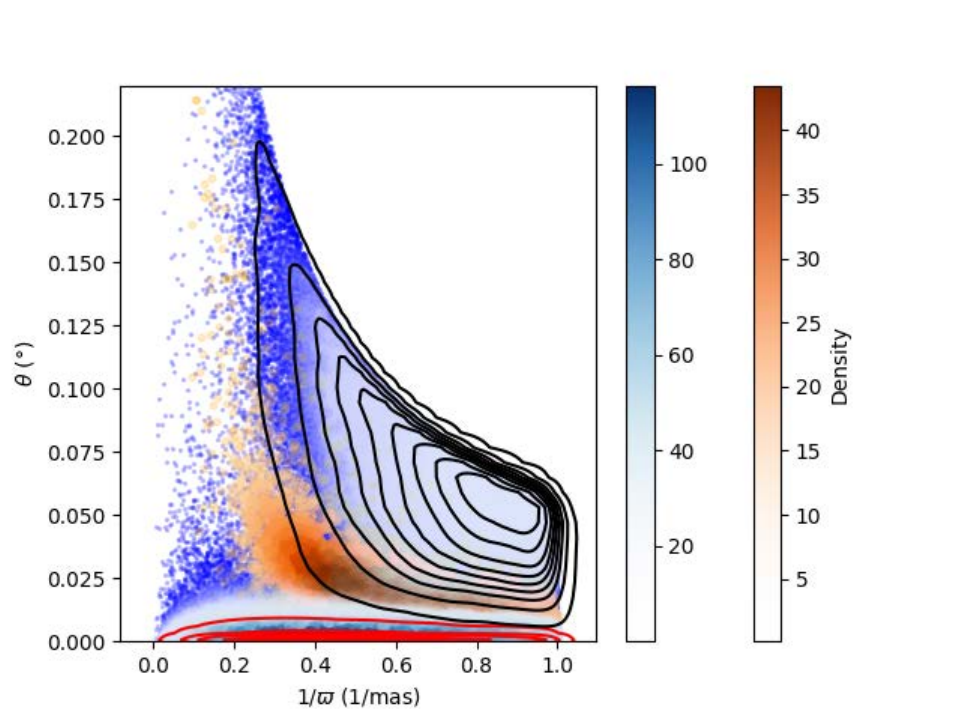}
    \caption{{The angular separations of classified binaries presented as a function of the inverse of the system's primary parallax. Both sets of misclassified (orange) and correctly classified (blue) binaries are overplotted by a Gaussian kernel density estimation. The binaries in the test set are denoted by the two contours, where the black and red contours correspond to the visual and true binary classes, respectively.}}
    \label{fig:winversetheta}
\end{figure}

From the ROC curve in Figure \ref{fig:perffig}, the ANN achieves an AUC score of 0.999, indicating that the model provides incredibly high quality classifications with respect to the TPR and FPR. The high classification quality of the ANN is also shown in the confusion matrix, where nearly every input binary in the test set is correctly classified as either a true binary or a visual binary. The model achieves high classification scores, with an accuracy of 0.993, a precision of 0.988, and a recall of 0.991.

\subsection{Computational Performance} \label{sec:compperf}
For the training, validation, and testing process, a standard laptop computer processor unit (CPU) was utilized. The 2.6 GHz 6-Core Intel Core i7 CPU took 8 minutes to execute the entire training and validation phases, computing 112 epochs at 4-5 s per epoch. The classification of the test set (313974 binaries) took 10 s.

\subsection{Misclassified Binaries} \label{sec:misclass}

In this section, the misjudged astrometric binary candidates are considered. Figure \ref{fig:misclasscomp} presents a histogram of the misclassified binaries and the correctly classified binaries from the test set as a function of their physical separation, which is effectively a plot of the confusion matrix in Figure \ref{fig:perffig} against the physical separation feature. This histogram verifies that there is notable influence given by the physical separation feature on the dataset. {Figure \ref{fig:winversetheta} plots the angular separations of the classified binaries against the inverse of the corresponding primary component's parallax and provides contours of the unclassified true and visual binary classes in this space. The misclassified binaries have mean angular separations of 0.021° ± 0.01°, a similar range of angular separations found in Figure \ref{fig:thetavs} where the intersection between the two classes also occurs.} 

{Interestingly, the majority of misclassified binaries fall under the visual binary class. This can be attributed to the dense clustering of the true binary class around the bottom edge of the model's estimated boundary between the two classes in the space found in Figure \ref{fig:winversetheta}. Furthermore, the rate at which misclassified binaries occur follows the general rate of change of the outer boundary of the visual binary class due to the dense clustering of the true binaries. As a result, there is a significantly greater number of false positives than false negatives.}

To solve this dilemma, additional features should be considered that can provide the ANN with the information necessary to correctly classify astrometric binaries when the other parameters are unhelpful. In particular, the incorporation of the RV feature may significantly increase model performance with the release of \textit{Gaia} DR4 (\citet{2023A&A...674A...1G}). While additional features may not always bring an increase in model performance, the selection of meaningful features, which RVs certainly are, is important for the maximization of model classification scores and can minimize the influence of the physical separation feature (\citet{kolleroptim}; \citet{KOHAVI1997273}). {Another alternative is to incorporate active learning (AL) into the modelling procedure of the ANN. AL is a technique in ML designed to optimize model performance by intelligently selecting points of interest to build a custom training data set for learning algorithms (\citet{settlesAL}). When model performance is low in a region of classification, the AL algorithm can suggest unlabelled data that would be most helpful for the learning algorithm (\citet{1996cs3104C}). The algorithm then makes queries to an oracle to determine the label for the suggested data and adds this data back into the training set to re-train the model. Considering that AL is especially useful in situations where the number of samples to label is large, an AL approach to select the relevant candidate binaries for training is a promising strategy for increasing model performance.}

\subsection{{Machine Learning Model Comparisons}} \label{sec:modelcomps}

{Given the strong definition of the true and visual binary classes by the physical separation feature, it is important to rigorously verify that the classification process is not defined by this feature. In comparison to the proposed ANN, a logistic regression (LR) model and an additional DNN, ANN-F4, are constructed. The LR model only considers the physical separation feature and ANN-F4 considers the four astrometric parameters (the proper motions and the parallaxes of the component stars) for training. Here, ANN-F4 has the same model architecture as the proposed ANN found in Figure \ref{fig:architpic}. Additionally, a cutoff point in the log(\textit{s} au$^{-1}$) space is set at $P_c = 4.61$, the mean of the overlap between the true and visual binary classes found in Figure \ref{fig:physepnetdata}, to compare the proposed ANN's classification scores at a predetermined physical separation threshold.} 

\begin{figure}[t]
    \centering
    \includegraphics[width=1.0\linewidth]{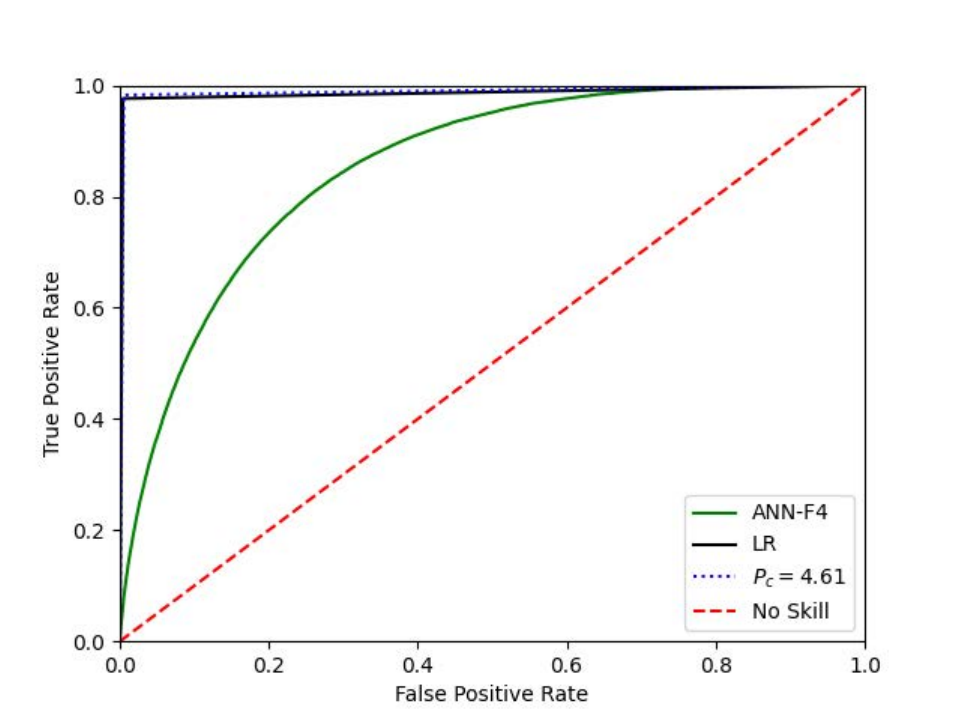}
    \caption{{ROC curves for each of the comparison models are plotted.}}
    \label{fig:DNN-LR}
\end{figure}

{In Figure \ref{fig:DNN-LR}, the associated ROC curves for the $P_c$ cut and the ANN-F4 and LR models are presented. When comparing classification models, it is usually best to consider the AUC of each model (\citet{LINGAUCACC}). To compare these comparison models to the proposed ANN, the DeLong test is utilized to provide the statistical difference between the AUCs of these models and the AUC of the proposed ANN (\citet{delong}). The DeLong test determines whether a statistically significant (\textit{p} $<$ 0.05) difference is present in the classification ability of the models. If a statistically significant difference is found, then it is taken that the AUC of the proposed ANN is indeed independent of the AUC of the comparison model and is thus a better classifier. The AUC, accuracy, precision, and DeLong \textit{p}-value of the $P_c$ cut and the ML models are shown in Table \ref{tab3}. The number of astrometric binaries in each test data set, \textit{n}, are also shown in this table.}

\begin{table}[t]
\setlength{\tabcolsep}{4.8pt}
\centering
\caption{{Statistical performance of the comparison models}}
\renewcommand{\arraystretch}{1.3}
\scriptsize
\begin{tabular}{l*{5}cc}
\hline
\hline
$$Model$$ & $$\textit{n}$$ & $$AUC$$ & $$Accuracy$$ & $$Precision$$ & $$\textit{p}-value$$\\
\hline
LR & 313794 & 0.989 & 0.991 & 0.988 & $$\textit{p} $<$ 0.05$$\\
ANN-F4 & 313794 & 0.854 & 0.817 & 0.745 &  $$\textit{p} $<$ 0.001$$\\
$P_c = 4.61$ & 1568967 & 0.988 & 0.991 & 0.986 & $$\textit{p} $<$ 0.01$$\\
\hline
\end{tabular}
\label{tab3}
\end{table}

{From the results shown in Table \ref{tab3}, these models do not perform as well as the proposed ANN. The expectation that the proposed ANN is the optimal classifier with regard to an AUC of 0.999 is verified by the DeLong test, where statistically significant tests are found for each ML model and the $P_c$ cut. With a statistically significant test and a lower accuracy score found in the LR model and the $P_c$ cut, it is evident that the proposed ANN is not solely defined by the physical separation feature. Instead, the additional information provided by the parallax and proper motion features permits for the proposed ANN to perform much better than these other models. Therefore, the statistical performance of the ML comparison models and the $P_c$ cut demonstrate that the selected features of the proposed ANN are necessary for optimal performance. It is important to note that the very statistically significant result (\textit{p} $<$ 0.001) found in ANN-F4 is a product of the model not being trained on the physical separation feature. Additionally, the difference in the statistical significance of the LR model and the $P_c$ cut despite both utilizing the same feature can be attributed to the shape of each test set relative to the shape of the test set of the proposed ANN.} 

{A final point of interest is the result shown from ANN-F4, which achieved quite high AUC and accuracy scores. While ANN-F4 clearly does not perform as well as the proposed ANN, this model does provide significant promise for the potential to classify astrometric binaries with only their observable astrometric parameters. Thus it is reasonable to suspect that the performance of ANN-F4 may increase with the widespread introduction of the RV feature that comes with the release of \textit{Gaia} DR4.}

\section{Conclusions} \label{sec:conclusions}

{In this paper, an artificial neural network (ANN) is proposed to classify a set of stars as either a true or visual binary based on their astrometric parameters (positions, proper motions, and parallaxes). To take advantage of the \textit{Gaia} mission, the learning algorithm was trained on a data set comprised of an extensive number of binary samples and astrometric parameters (positions, proper motions, and parallaxes) from \textit{Gaia} DR3. After careful testing, the ANN performs exceptionally well on withheld data, achieving very high scores in every classification metric, and is not subject to overfitting.}

{It was observed that the constructed data set was strongly defined by the physical separation feature. As an experiment, a logistic regression model, an additional neural network, and a predetermined cut in the physical separation space were created to compare to the proposed ANN. The comparison models were each outperformed by the ANN, implying that every selected feature is necessary for optimal model performance. In addition, this experiment demonstrated that there is significant potential for classifying astrometric binaries with regard to only their observable astrometric parameters. Therefore, it is suspected that it will become feasible to classify astrometric binaries and neglect the consideration of the physical separation and angular separation features with the robust introduction of radial velocities (RVs) when \textit{Gaia} DR4 is published.}

Compared to current classification techniques, the proposed ANN offers significant benefits for classifying astrometric binaries by saving significant net computing time and costs. Run on a cheap and low performance CPU, the training and validation phases of the ANN took about 8 minutes, while the classification of the test set (approximately 314000 binaries) took only 10 s. If the developed ANN was run on a similar CPU or graphics processing unit (GPU) compared to other machine learning (ML) models (e.g., \citet{2020ApJ...897L..12S}), the above computing times would be significantly lower.

As various sky surveys continue to provide ample data sets of binary star candidates, astronomers need tools to identify astrometric binaries in an effective and efficient way. In particular, the \textit{Gaia} mission has proven to be difficult to analyze in depth due to the sheer volume of data obtained for nearly two billion point sources. Similarly, future sky surveys that provide ample astrometric data sets require an automation tool to be fully taken advantage of. Therefore, the proposed ANN is a viable method for the analysis of large databases, thereby providing a means for astronomers to take full advantage of the \textit{Gaia} mission and future sky surveys for other and similar research purposes.

\section*{Data Availability}
The \textit{Gaia} DR3 database can be found in the \textit{Gaia} archive at \href{https://gea.esac.esa.int/archive}{https://gea.esac.esa.int/archive}. The code for the proposed ANN can be found at \href{https://zenodo.org/records/13621762}{https://zenodo.org/records/13621762}.
\section*{Acknowledgements}
Gratitude is expressed to Pat Boyce for his continued support of this work during the research process and the writing of this manuscript.

This work has made use of data from the European Space Agency (ESA) mission \textit{Gaia} (\href{https://www.cosmos.esa.int/gaia}{https://www.cosmos.esa.int/gaia}), processed by the \textit{Gaia} Data Processing and Analysis Consortium (DPAC, \href{https://www.cosmos.esa.int/web/gaia/dpac/consortium}{https://www.cosmos.esa.int/web/gaia/dpac/consortium}). Funding for the DPAC has been provided by national institutions, in particular the institutions participating in the \textit{Gaia} Multilateral Agreement.

\textit{Software}: Python (\citet{2011CSE....13b..22V}), Numpy (\citet{2011CSE....13b..22V}), Pandas (\citet{mckinney-proc-scipy-2010}), Scikit-learn (\citet{scikit-learn}), Tensorflow (\citet{tensorflow2015-whitepaper}), Keras (\citet{chollet2015keras}), Seaborn (\citet{2021JOSS....6.3021W}), Matplotlib (\citet{2007CSE.....9...90H}).

\section*{ORCID iDs}
Joe Smith \orcidlink{0009-0008-3226-6205} \href{https://orcid.org/0009-0008-3226-6205}{https://orcid.org/0009-0008-3226-6205}

\bibliography{citations}{}
\bibliographystyle{aasjournal}

\end{document}